\documentstyle[aps,amsfonts,amssymb,prb,epsfig,floats]{revtex}
\begin{document}
\draft \preprint{}
\twocolumn[\hsize\textwidth\columnwidth\hsize\csname
@twocolumnfalse\endcsname
\title{Electron-Like Fermi Surface and Remnant ($\pi$,0) Feature in Overdoped La$_{1.78}$Sr$_{0.22}$CuO$_4$}
\author{T. Yoshida$^1$, X. J. Zhou$^2$, M. Nakamura$^3$, S. A. Kellar$^2$, P. V. Bogdanov$^2$, E. D. Lu$^4$, A. Lanzara$^{2,4}$,  Z. Hussain$^4$, \protect\newline A. Ino$^5$, T. Mizokawa$^6$, A. Fujimori$^{1,6}$, H. Eisaki$^2$, C. Kim$^2$, Z.-X. Shen$^2$, T. Kakeshita$^7$, S. Uchida$^7$}
\address{$^1$Department of Physics, University of Tokyo, Bunkyo-ku,
  Tokyo 113-0033, Japan}
\address{$^2$Department of Applied Physics and Stanford Synchrotron Radiation Laboratory, Stanford University, Stanford, CA94305}
\address{$^3$Department of Physics, Nara University of Education, Takabatake-cho, Nara 630-8523, Japan}
\address{$^4$Advanced Light Source, Lawrence Berkeley National Lab, Berkeley, CA 94720}
\address{$^5$Synchrotron Radiation Research Center, Japan Atomic Energy Research Institute, SPring-8, Mikazuki, Sayo, Hyogo 679-5198, Japan}
\address{$^6$Department of Complexity Science and Engineering, University of Tokyo, Bunkyo-ku, Tokyo 113-0033,
  Japan}
\address{$^7$Department of Advanced Materials Science, University of Tokyo, Bunkyo-ku, Tokyo 113-8656, Japan}
\date{\today}
\maketitle
\begin{abstract}                
We have performed an angle-resolved photoemission study of
overdoped La$_{1.78}$Sr$_{0.22}$CuO$_4$, and have observed sharp
nodal quasiparticle peaks in the second Brillouin zone that are
comparable to data from Bi$_2$Sr$_2$CaCu$_2$O$_{8+\delta}$. The
data analysis using energy distribution curves, momentum
distribution curves and intensity maps all show evidence of an
electron-like Fermi surface, which is well explained by band
structure calculations. Evidence for many-body effects are also
found in the substantial spectral weight remaining below the Fermi
level around ($\pi$,0), where the band is predicted to lie above
E$_F$.    \end{abstract} \pacs{PACS numbers: 74.25.Jb, 71.18.+y,
74.72.Dn, 79.60.Bm}
\vskip2pc]
%

\narrowtext

Studies of low lying excitations and the Fermi surface in the high
temperature superconductors by angle resolved photoemission
spectroscopy (ARPES) have mostly been focused on
Bi$_2$Sr$_2$CaCu$_2$O$_{8+\delta}$
(BSCCO)\cite{chuang,feng,bogdanov,gromko,fretwell,fink} and
YBa$_2$Cu$_3$O$_{7-y}$ (YBCO)\cite{Matthias}. In these systems,
however, additional features derived from complicated crystal
structures, such as the Bi-O superstructures in BSCCO and the Cu-O
chains in YBCO, have made the analysis rather complicated. In
particular, the Bi-O superstructure complicates the interpretation
of spectra around ($\pi$,0), resulting in the controversy over the
Fermi surface
geometry\cite{chuang,feng,bogdanov,gromko,fretwell,fink}.
La$_{2-x}$Sr$_{x}$CuO$_4$ (LSCO), by virtue of the absence of
these effects and the availability of high quality single crystal
samples over the entire doping range, provides the opportunity to
advance our understanding of the high temperature superconductors.
In the underdoped regime, earlier studies have uncovered the
presence of two electronic components\cite{ino1,ino2,ino3}, a
systematic suppression of the spectral weight near ($\pi/2,\pi/2$)
(when compared with that of overdoped samples or BSCCO for data
taken under the same conditions), and straight Fermi surface
segments near ($\pi$,0) of width $\sim \pi/2$\cite{zhou1,zhou2},
which have been interpreted as evidence for electronic
inhomogeneities. In the overdoped regime, an electron-like Fermi
surface was observed in the high-T$_c$ superconductors for the
first time\cite{ino1}, but despite this progress, important
problems remain. Since ARPES data from underdoped samples are very
broad, worries about the sample quality persist. There are also
questions about the effects of the photoemission matrix
element\cite{bansil}, which makes it difficult to extract
quantitative information about stripe effects on nodal spectral
weight by comparing the experiments and theoretical calculations
that predict suppression\cite{tohyama,fleck,machida}. We address
these important questions by performing a detailed study of
overdoped LSCO ($x$= 0.22) where the stripe effects are expected
to be weak, and have the aid of reliable band structure
calculations (unlike the case of BSCCO). We performed ARPES in
three Brillouin zones (BZ) and performed numerical simulations to
investigate the matrix element effects. In the second BZ, using a
favorable polarization, we have identified a sharp spectral
feature along the diagonal direction in this sample that is
comparable to that of BSCCO. This observation demonstrates that
the surface quality of LSCO is comparable to that of BSCCO and
gives credence to LSCO data in the underdoped region, since the
sample quality of LSCO in the underdoped region is expected to be
improved with decreasing Sr content. Our detailed analysis shows
that the compound has an electron-like Fermi surface with
additional spectral weight near ($\pi$,0), which is the remnant of
the flat band feature in the underdoped samples. The detailed
comparison of experiment and simulation also provides a better
understanding of the matrix element effect in these experiments.


The ARPES measurements were carried out at BL10.0.1.1 of the
Advanced Light Source, using incident photons with an energy of
55.5 eV. We used a SCIENTA SES-200 spectrometer in angle mode,
where one can collect spectra over $\sim$14 degrees, corresponding
to a momentum width of $\sim$1.1$\pi$ (in units of $1/a$, where
$a\sim 3.8\mathrm{\AA}$ is the lattice constant). The total energy
and momentum resolution was about 20 meV and 0.02$\pi$,
respectively. We studied high quality single crystals of LSCO with
$x= 0.22$ grown by the traveling-solvent floating-zone method. The
measurements were performed in an ultra high vacuum of 10$^{-11}$
Torr at 20 K and the samples were cleaved {\it in situ}. The
position of the Fermi level ($E_F$) was calibrated with gold
spectra.

\begin{figure}[!t]
\centerline{\epsfig{figure=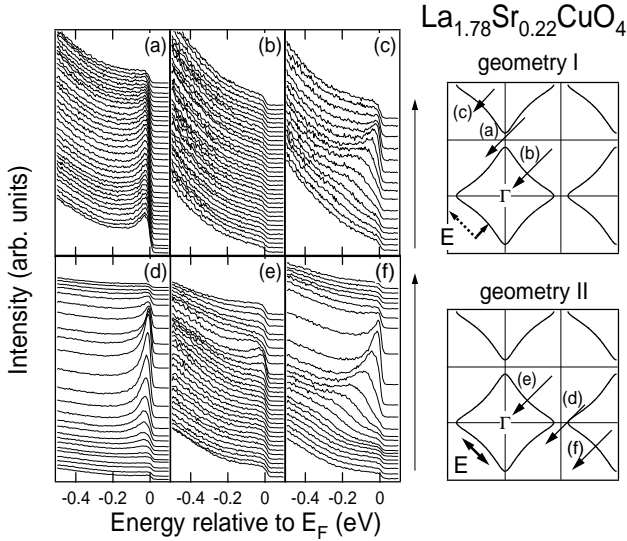,width=1.0\linewidth,clip=}}\vspace{0.5pc}
\caption{Energy distribution curves for
La$_{1.78}$Sr$_{0.22}$CuO$_4$ at 20 K. Right panels show the
momentum space for each cut and the E-vector in the two
geometries. For details, see the text.} \label{EDC}
\end{figure}


Figure \ref{EDC} shows energy distribution curves (EDC's) for two
different experimental geometries along the directions shown in
the right panel. In both geometries, the light is incident to the
surface with an incident angle of $\sim 7^\circ$. The EDC's in
geometry I were measured by rotating the sample, which causes the
changes in the in-plane $E$-vector of the incident photons as
shown in the right panel. On the other hand, in geometry II, the
spectra were measured by moving the analyzer, and therefore the
in-plane $E$-vector is fixed. While in geometry I, the $E$-vector
was almost normal to the sample surface, resulting in a small
in-plane $E$ component, in the geometry II, the $E$-vector was
almost parallel to the sample surface. Panels (a) and (d) show
ARPES spectra around the $(\pi,0)$ point. These spectra show a
single peak with little dispersion, which will be addressed in the
discussion below as a remnant of the ``flat band" feature
appearing below $E_F$ even for $x<0.2$\cite{ino1}. Panels (b) and
(e) give the nodal cut [(0,0) to ($\pi,\pi$) direction] in the
first BZ for each geometry. One can see that panel (e) shows a
dispersive feature while panel (b) shows almost no corresponding
spectral feature, due to transition matrix element effects. Thus,
compared to the previous results, we could observe the nodal
feature in the first BZ more clearly in geometry II. Furthermore,
as shown in panels (c) and (f), we found that the nodal state
feature become stronger in the second BZ, particularly in geometry
II. As a whole, geometry II gives clearer dispersive features,
probably because the in-plane $E$-vector is much larger than that
in geometry I. These results show that there are significant
matrix element effects which should be carefully considered in
order to extract intrinsic information. It should be emphasized
that the intensity of the first BZ nodal state in underdoped LSCO
is still much weaker than that in BSCCO or overdoped LSCO observed
under the same experimental conditions. These relative changes in
data with doping under the same experimental geometry are
intrinsic but the matrix element effect makes quantitative
analysis more difficult.

One may suspect that the weak dispersion of the features in
underdoped LSCO, particularly in the nodal direction \cite{ino2},
may be due to inferior surface quality (roughness, defects etc.).
The present results have shown that the sharpness of the nodal
feature seen in geometry II is comparable to that in BSCCO, which
demonstrates the high quality of the LSCO surfaces. One can expect
that the sample quality decreases with increasing $x$ because of
the disorder introduced by Sr and the difficulty in the crystal
growth due to the Sr solubility limit. In fact, underdoped samples
give better cleavage than overdoped samples, which possibly
indicates the better surface quality for the underdoped samples.
Therefore, the ARPES results in the underdoped region, which show
broad features around ($\pi$, 0) and interpreted as a
two-component feature\cite{ino2}, should be reliable results.

\begin{figure}[!b]
\centerline{\epsfig{figure=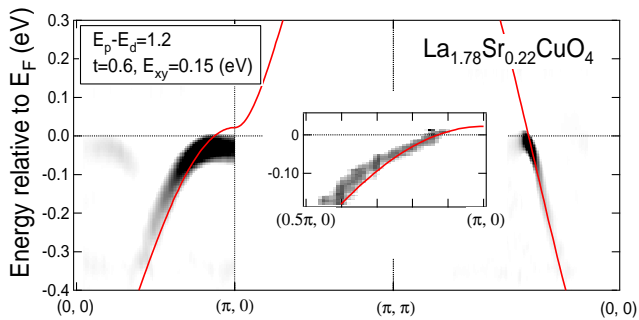,width=1.0\linewidth,clip=}}\vspace{0.5pc}
\caption{Energy dispersion and a tight-binding fit for
La$_{1.78}$Sr$_{0.22}$CuO$_4$. The gray plot in the inset was
obtained by normalizing the spectra to the peak in each MDC.}
\label{dispersion}
\end{figure}

The band dispersions along the (0,0) to ($\pi,0$) and (0,0) to
($\pi,\pi$) directions, which have been derived from the ARPES
spectra by taking the second derivatives, are shown in Fig.
\ref{dispersion}. In going from (0,0) to ($\pi$,0), the band
reaches $E_F$ at $k_x \sim 0.8 \pi$ but substantial spectral
weight remains below $E_F$ up to ($\pi,0$). However, as shown in
the inset, the peak in the momentum distribution curves (MDC's)
clearly crosses the Fermi level around ($0.85\pi,0$) concomitant
with the decrease of spectral weight for $k_x > 0.85\pi$. This
behavior is not seen in optimally-doped and underdoped LSCO along
the (0,0)-($\pi$,0) line \cite{ino1}. The LDA band calculation
also predicts an electron like Fermi surface for $x\geq 0.17$
\cite{xu,pickett}, although the calculation shows finite band
dispersions along the c-axis and the Fermi surface is somewhat
$k_z$-dependent. The $k_z$ dispersion is expected to be strongly
renormalized in real materials\cite{uchida}. Therefore, we state
that the $x=0.22$ has an electron-like Fermi surface centered at
(0,0). The red line shows a tight-binding fit to the experimental
band dispersion with parameters $E_p-E_d$= 1.2 eV, $t_{pd\sigma}$=
0.5 eV, $E_{xy}\equiv t_{pp\sigma}/2-t_{pp\pi}/2$=0.15 eV. Here,
$t_{pd\sigma}$ and $E_{xy}$ are the transfer integrals for the
nearest-neighbor O 2$p_\sigma$-Cu 3$d_{x^2-y^2}$ and O 2$p_x$-O
2$p_y$ overlap, respectively. The best fit values for
$t_{pd\sigma}$ and $E_{xy}$ are much smaller than those obtained
from a tight-binding fit to the LDA band-structure
calculation\cite{hybertsen} due to the band narrowing
corresponding to the mass enhancement by a factor of $\sim 3$ in
the overdoped region \cite{inoIPES}.

\begin{figure}[!t]
\centerline{\epsfig{figure=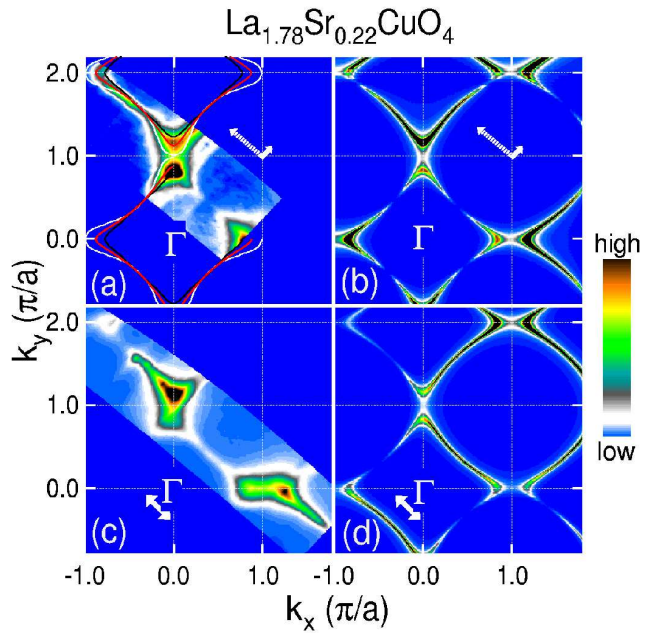,width=1.0\linewidth,clip=}}\vspace{0.5pc}
\caption{Spectral weight integrated within 30 meV of the Fermi
level. White arrows designate the E vector. (a)(c); Experiment.
(b)(d); Simulation. White and black curves in (a) represent the
Fermi surfaces of band calculation\protect\cite{xu} at $k_z=0$ and
$\pi/c$, respectively, and red curves represents the Fermi surface
from the present tight-binding fit. Note that the spectral weight
in the nodal direction is enhanced for geometry II [(c) and (d)]
compared to geometry I [(a) and (b)].} \label{nk}
\end{figure}

Figure \ref{nk} shows the spectral weight plot integrated over a
30 meV window around the Fermi level [(a),(c)] for both geometries
and simulations of the spectral weight plot for each geometry
including transition matrix element effects [(b),(d)] (see below).
Integrating over the narrow window of the order of the energy
resolution makes it possible to obtain the spectral weight at
$E_F$, which approximately represents the Fermi surface. The Fermi
surfaces at $k_z=0$ and $\pi/c$ from the band-structure
calculation for La$_{2-x}M_x$CuO$_4$ ($x= 0.2$) \cite{xu}, as well
as the Fermi surface from the present tight-binding fit are
superimposed on Fig. \ref{nk}(a). As a whole, they well describe
the global features related to the Fermi surface obtained from
ARPES. In particular, looking at the nodal direction in the second
BZ, they agree well with each other. The volume enclosed by the
tight-binding Fermi surface is $S_{\rm FS}\simeq0.8\times2\pi^2$,
which satisfies Luttinger's sum rule [$S_{FS}=2\pi^2(1-x)$] within
experimental accuracy.

Here, we have performed simulations of spectral weight
distribution including matrix-element effects by using the same
method described in \cite{daimon,matsushita}. The simulation
method which we apply here has given a good account of the
photoemission results on graphite \cite{daimon} and TaSe$_2$
\cite{matsushita}. Therefore, this is a good starting point to
understand the matrix element effect in the present ARPES results,
at least qualitatively. First, we consider the initial state $|i>=
(1/\sqrt{N})\sum_{i,j} e^{-iq\cdot (R_j+\tau_i)  }
a_i\phi_i(r-R_j-\tau_i) $ as the two-dimensional tight-binding
ground state, where $\phi_i$ is the atomic orbital Cu
3$d_{x^2-y^2}$ or O 2$p_x, p_y$. The matrix element factor is
given by $M(k)=|a_{d_{x^2-y^2}}A_{d_{x^2-y^2}}+a_{p_x}
A_{p_x}+a_{p_y} A_{p_y}|^2$, where $A_{d_{x^2-y^2}}, A_{p_y},
A_{p_x}$ are angular distribution factor from each atomic orbital
Cu 3$d_{x^2-y^2}$, and O 2$p_x, p_y$ \cite{goldberg}. $a_d,
a_{p_x}$, and $a_{p_y}$ have been determined from the
tight-binding fit to the dispersion of ARPES result as shown in
Fig. \ref{dispersion}. The spectral function $A({\bf k},\omega)$
of the tight-binding band was approximated by broadening the
$\delta$-function with width of $\omega^2$, where $\omega$ is in
units of eV. Then, the simulated momentum distribution of spectral
weight is given by $n({\bf k})=M({\bf k}) \int I({\bf
k},\omega)d\omega$, where $I({\bf k},\omega)$ is spectral weight
broadened by energy resolution of 20 meV.

As shown in Fig. \ref{nk}(b), in geometry I, the nodal states in
the first BZ show almost no spectral weight while those in second
BZ are enhanced, consistent with the experimental data in Fig.
\ref{nk}(a). This implies that the suppression in geometry I
compared to geometry II is caused by matrix element effects.
According to the simulation, the smaller matrix elements in the
first BZ are a result of the combination of the symmetry of the
$d_{x^2-y^2}$ orbital and the large out of plane component in the
$E$-vector. The enhancement in the second BZ may be caused by the
angular distribution factor of the three atomic orbitals, because
they have a small emission probability for small angles when the
$E$-vector is vertical to the surface. In geometry II, we can see
the clear dispersion and spectral weight in the nodal state in the
first BZ as shown in Fig. 1(e) and Fig. 3(c). The simulation in
Fig. 3(d) shows an enhancement of the spectral weight in the
(0,0)-($\pi,\pi$) direction compared to the nodal states of the
first BZ in geometry I, which qualitatively agrees with the
experimental results.

\begin{figure}[!t]
\centerline{\epsfig{figure=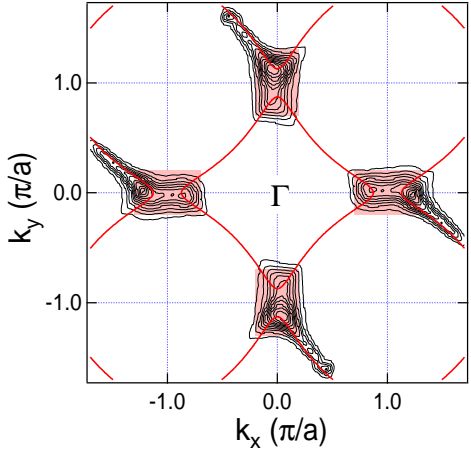,width=0.85\linewidth,clip=}}\vspace{0.5pc}
\caption{Contour plot of the spectral weight shown in Fig. 3(c)
symmetrized with respect to the $\Gamma$ point. Red curves show
the electron-like Fermi surface obtained from the fit to the
experiment as shown in Fig. 2. Shaded regions around ($\pi$,0)
reflect a remnant of the ``flat band".} \label{FS}
\end{figure}


While the overall features in the experimental results agree with
the simulation as shown above, there was still a discrepancy
between them regarding the spectral weight distribution near
($\pi,0$). Here, we discuss the possible origin of the strong
spectral weight around the ($\pi,0$) point which cannot be
described within the simple two dimensional electron-like Fermi
surface picture. As shown in Fig. 4, the spectral weight
distribution around ($\pi,0$) shows a relatively straight contour
along the $k_x$ direction, which is slightly narrower than that in
Nd-LSCO ($|k_y|<\pi/4$)\cite{zhou1}. This spectral weight
distribution is very similar to that of the flat band which
appears below $E_F$ for smaller $x$. Presumably the ($\pi,0$) flat
band feature which exists in the optimum and underdoped regions
does not completely lose its spectral weight even for $x= 0.22$
(and probably for $x=0.3$, see data in Ref.\cite{ino1}) where the
saddle point is located above $E_F$. Therefore, the spectral
weight around ($\pi,0$) appears as a remnant of the ``flat band".
Recently, there have been debates as to whether there exists an
electron-like Fermi surface in BSCCO or not
\cite{chuang,feng,bogdanov,gromko,fretwell,fink}. While this
discussion has been complicated by the Bi-O superstructures, the
present results from overdoped LSCO clearly shows an electron-like
Fermi surface with much less ambiguity. One may suspect that the
matrix element may suppress spectral weight only around the
$(\pi,0)$ point making the hole-like Fermi surface appear
electron-like. However, as far as the matrix element simulation
using the simple tight-binding scheme is concerned, such a
suppression localized in momentum space around ($\pi$,0) is
difficult to explain. This is consistent with an earlier
calculation\cite{bansil}. In the case of BSCCO, the possibility of
additional states at ($\pi,0$) has been proposed \cite{chuang}. In
the present case, strong intensity around ($\pi,0$), which is not
predicted by the simulation, always exists irrespective of
polarization geometries, while the nodal states are strongly
affected by matrix element effects. This implies an intrinsic
unusual electronic structure such as stripes associated with the
``flat band" feature.

As seen in Fig. \ref{FS}, we have shown that the electronic
structure of LSCO with $x=0.22$ shows two features. One is the
nodal Fermi surface which is found to be consistent with the band
structure calculation. The other is the remnant flat band which
gives rise to a straight segment of spectral weight near the
($\pi,0$) region. These two features are qualitatively similar to
those observed in Nd-LSCO and LSCO with $x=0.15$ samples
\cite{zhou1,zhou2}, although quantitatively the flat band effect
in LSCO with $x=0.22$ is weaker. This dual nature of the
electronic structure can be explained in terms of order-disorder
stripes competition \cite{zhou2}. The origin of the nodal state
can be understood by considering strongly disordered stripes
and/or weakened stripe order. In the latter case, one may begin to
recover the underlying band structure. The underlying electronic
structure will manifest itself stronger as the charge ordering
effect becomes weaker in LSCO with $x=0.22$.


In conclusion, we have unambiguously observed an electron-like
Fermi surface in slightly overdoped LSCO with $x=0.22$, which
agrees with the band-structure calculation. By utilizing the
matrix element effects, we have obtained clear dispersion and
sharp peak features for the nodal states comparable to BSCCO. The
transition matrix element effects have been discussed by
simulations, which account for the enhancement in the second BZ
and for different geometries. Although the stripe effect in the
present sample is weaker, the observation of the nodal spectral
weight and the straight segment near ($\pi,0$) is consistent with
the picture of order-disorder competition of stripes in the
system.

This work was supported by a Grant-in-Aid for Scientific Research
``Novel Quantum Phenomena in Transition Metal Oxides", a Special
Coordination Fund from the Science and Technology Agency and the
New Energy and Industrial Technology Development Organization
(NEDO). Advanced Light Source of Lawrence Berkeley National
Laboratory is operated by U.S. Department of Energy's Office of
Basic Energy Science, Division of Material Science.

\end{document}